\documentclass[reprint,amsmath,amssymb,
aps,prl,twocolumn,superscriptaddress,groupedaddress]{revtex4}
\usepackage{graphicx}
\usepackage{dcolumn}
\usepackage{bm}
\usepackage{float}
\usepackage{amssymb}
\begin{document}

\title{Electron scattering in tantalum monoarsenide}

\author{Cheng-Long Zhang}
\affiliation{International Center for Quantum Materials, School of Physics, Peking University, Beijing 100871, China}

\author{Zhujun Yuan}
\affiliation{International Center for Quantum Materials, School of Physics, Peking University, Beijing 100871, China}

\author{Qing-Dong Jiang}
\affiliation{International Center for Quantum Materials, School of Physics, Peking University, Beijing 100871, China}

\author{Bingbing Tong}
\affiliation{International Center for Quantum Materials, School of Physics, Peking University, Beijing 100871, China}

\author{Chi Zhang}
\affiliation{International Center for Quantum Materials, School of Physics, Peking University, Beijing 100871, China}
\affiliation{Collaborative Innovation Center of Quantum Matter, Beijing 100871, China}

\author{X.C. Xie}
\affiliation{International Center for Quantum Materials, School of Physics, Peking University, Beijing 100871, China}
\affiliation{Collaborative Innovation Center of Quantum Matter, Beijing 100871, China}

\author{Shuang Jia\footnote{gwljiashuang@pku.edu.cn}}
\affiliation{International Center for Quantum Materials, School of Physics, Peking University, Beijing 100871, China}
\affiliation{Collaborative Innovation Center of Quantum Matter, Beijing 100871, China}

\date{\today}

\begin{abstract}
We report comprehensive studies of the single crystal growth and electrical transport properties for various samples of TaAs, the first experimental confirmed inversion symmetry-breaking Weyl semimetal.
The transport parameters for different samples are obtained through the fitting of two-band model and the analysis of Shubnikov de-Haas oscillations. We find that the ratio factor of transport lifetime and quantum lifetime is intensively enhanced when the Fermi level approaches the Weyl node. This result is consistent with the side-jump interpretation derived from a chirality-protected shift in the scattering process for a Weyl semimetal.

\end{abstract}

\maketitle

\section{\romannumeral1. INTRODUCTION}
A semimetal has a small overlap between the bottom of the conduction band and the top of the valence band. The small occupation of the Fermi surface (FS) in a semimetal can cause some unique electrical transport features unlike those in typical metals. For instance, the compensation of the electrons and holes at the Fermi level (FL) in a semimetal leads to a resonance in the charge transport \cite{ali_WTe2_2014}. Large magnetoresistance ($\mathrm{MR} \equiv \Delta \rho /\rho _{H=0}$) has been observed in several compensated semimetals \cite{ali_WTe2_2014,Reentrant,GMRbi}, including bismuth, which is an archetypical semimetal: high-purity bismuth has a carrier concentration as low as $\sim$$10^{17}$ $\mathrm{cm^{-3}}$ and its FS occupies a mere $10^{-5}$ of the Brillouin zone. Bismuth can serve as a model system for studying the  dilute Dirac electron gas \cite{behnia2007signatures,li2008phase}. Discovery of a new topological semimetal with different band structures and electron characteristics has been one of the central task for the condensed matter physics society. The newly discovered topological semimetals include Dirac semimetals Cd$_3$As$_2$ \cite{Cd3As2_theory} , Na$_3$Bi \cite{Na3Bi_theory} and the Weyl semimetal TaAs isostructural family  \cite{TaAs_Arpes_Hasan,TaAs_Arpes_Hongding,TaAs_Chenyulin,NbAs_Arpes_Hasan,TaP_Dinghong,TaP_Suyang,TaAs_theory_Hsin,TaAs_theory_Xidai}. The TaAs family exhibit many interesting transport properties that are attributed to their exotic low-energy excitations. These properties include negative longitudinal MR induced by the chiral anomaly \cite{TaAs_ABJ_SJ,TaAs_transport_GengfuChen,NbAs_Chiral_Xuzhuan,TaP_MiaosiqiChiral,TaP_chiralMing}, extremely large linear MR and ultrahigh carrier mobilities \cite{TaP_zcl,TaP_Claudio,NbAs_Luoyongkang,NbP_transport_Claudia}.
However when we compare the results of the transport experiments on each compound of the TaAs family, we found that the properties of different samples are indeed very different. Understanding the mechanism underneath the sample difference should be very important for addressing the Weyl quasiparticles. Unfortunately, researchers usually only present the data from their `best' samples in one report.

We here present our results in an extensive manner after a systematic exploration on the methods of sample growth, electrical measurements on different samples and comprehensive analyses on transport parameters. We show that the growth procedures significantly affect the sample's quality and transport properties for TaAs. The position of the FL is one of the most important factors to affect the scattering procedures. The large lifetime ratio factor at low temperatures indicates that the large mobility comes from the strong backscattering protection in TaAs. These exotic properties promise TaAs as a platform for investigating the topological electrons.


\section{\romannumeral2. materials and methods}

Previous study reported that the single crystals of TaAs can be grown via a standard chemical vapor transfer (CVT) method \cite{ MurrayTA_mainGrowth}. However we found that the yield and quality of the single crystals are not stable in different batches when we followed the procedure in Ref. \cite{MurrayTA_mainGrowth}. Therefore we optimized the CVT method by attempting different temperature gradients and agents. Below we describe the modified growth procedures in detail.

Polycrystalline samples were prepared from stoichiometric mixtures of Ta (99.98\%) and As (99.999\%) powders in an evacuated quartz ampoule at 1473 K. Then we grew the single crystals via the CVT method with a positive temperature gradient similar as the procedure in Ref. \cite{MurrayTA_mainGrowth}. The powder of TaAs (300 mg) and the transport agent (I$_2$, 10 $\mathrm{mg/cm^3}$ or SnI$_4$, 30-50 mg) were sealed in a 30 cm-long evacuated quartz ampoules. The end of the sealed ampoule with charges was placed horizontally at the center of a single-zone tube furnace. The center was slowly heated up to 1273 K and kept the temperature for 5 days. Small crystals about 0.5 mm in scale were obtained at the cold end which was about 973 K during the growth. The yield is not stable from batch to batch in this procedure.

We found that the CVT growth with reversed temperature gradients and different transport agents can produce large crystals in a more reproducible manner. Polycrystalline TaAs (300 mg) and one of the transport agents (SnI$_4$, NbI$_5$, TeI$_4$, TaBr$_5$, TaCl$_5$ or BiBr$_3$; 20-40 mg) were sealed in  23 cm-long evacuated quartz ampoules. The ampoules were placed in a three-zone furnace with the end containing the charges at the center of the central zone while another end at the center of the side zone. The central and side zones were slowly heated up to 1073 K and 1273 K, respectively, and then kept these temperatures for 8 days.

The procedure described above yields different shapes of single crystals like pyramids or flat blocks (Fig. 2(a)). All the samples can be seen as part of truncated octahedron indeed. X-ray diffraction measurements confirmed that the square surfaces on the crystals are the crystallographic $\bf{c}$ planes with four-fold rotational axes. The size of the crystals is about 0.5-2 mm in general, but bigger crystals can be obtained by extending the growing period to one month.

All the measurements were preformed on the polished  $\bf{c}$ planes with the electric current passing along the $\bf{a}$ direction in this paper. The crystal growth conditions, residual resistance ratio (RRR) and the MR at 2 K and 9 T of the samples  are summarized in Table \uppercase\expandafter{\romannumeral1}.
We empirically found that the samples with golden metallic lustre have the highest RRR and mobility among all the samples. We also found that some large crystals ($>$ 1 mm) have many stripes on their side faces, which are likely due to the stacking disorders along c-axis \cite{disorder}. These samples in general have much smaller RRR values. It has been reported that the pnictide deficiency induces stacking disorders in the isostructural compounds \cite{transposition_TaAs_1963}. According to our observation, the stacking disorders affect the quality of large crystals seriously, which is an obstacle for obtaining large high-quality crystals.

The values of MR and RRR for various samples of TaAs were shown in a double-logarithmic plot in Fig. 1. The MR at 2 K and 9 T changes from 10$^4$ to less than 10 for the samples grown with different agents. It is noteworthy that the MR at 2 K seems to follow the power-law of MR $\propto$ RRR$^{2.4}$. Such power law was reported in WTe$_2$ as well \cite{WTE2RRR}.

\begin{table}[h!]
\begin{flushleft}
\caption{\label{table1} Summarized growth conditions, residual resistance ratio (RRR) and magnetoresistance (MR) at 2 K and 9 T for representative samples of TaAs. All the growth started from 250 mg of TaAs powders and one of the agents listed below. }
\begin{ruledtabular}
\begin{tabular}[t]{llllll}
$Sample\#$  & $Agent$(mass) & $T$ & $l_{tube}$ & $RRR$ & $MR$(9T) \\
 & mg & $^o$$C$ & cm & $\frac{\rho_{xx}(300 \mathrm {K})}{\rho_{xx}(2 \mathrm {K})}$ & $\Delta\rho/\rho_0$  \\
\colrule
D1 & SnI$_4$(60) & 800$\rightarrow$1000 & 23 & 25 & 1329\\
D2 & TeI$_4$(60) & 1000(1 zone) & 23 & 6 & 81\\
D3 & TeI$_4$(70) & 1000(1 zone) & 23 & 5 & 86\\
D5 & NbI$_5$(50) & 700$\rightarrow$900 & 23 & 2 & 3\\
D6 & SnI$_4$(60) & 800$\rightarrow$1000 & 23 & 11 & 840\\
D10 & SnI$_4$(60) & 930$\rightarrow$1000 & 23 & 22 & 2473\\
D11 & SnI$_4$(60) & 850$\rightarrow$930 & 23 & 38 & 5143\\
B4 & TaBr$_5$(150) & 970$\rightarrow$830 & 23 & 7 & 293\\
B5 & TaCl$_5$(120) & 970$\rightarrow$830 & 23 & 8 & 261\\
B6 & TaCl$_5$(170) & 970$\rightarrow$830 & 23 & 7 & 226\\
B7 & TaBr$_5$(220) & 970$\rightarrow$830 & 23 & 41 & 3180\\
B8 & TaCl$_5$(120) & 970$\rightarrow$830 & 23 & 11 & 308\\
B9 & TaCl$_5$(170) & 970$\rightarrow$830 & 23 & 13 & 423\\
S1 & SnI$_4$(60) & 1000(1 zone) & 30 & 49 & 5449 \\
S3 & SnI$_4$(80) & 800$\rightarrow$1000 & 24 & 12 & 873\\
S4 & SnI$_4$(80) & 800$\rightarrow$1000 & 24 & 17 & 1476\\
S5 & I$_2$(50) & 1000(1 zone) & 30 & 5 & 4\\
S9 & SnI$_4$(100) & 800$\rightarrow$1000 & 24 & 21 & 1593\\
S14 & SnI$_4$(100) & 800$\rightarrow$1000 & 24 & N/A & 1244\\
\end{tabular}
\end{ruledtabular}
\end{flushleft}
\end{table}

\begin{figure}[h!]
  \begin{center}
  \includegraphics[clip, width=0.45\textwidth]{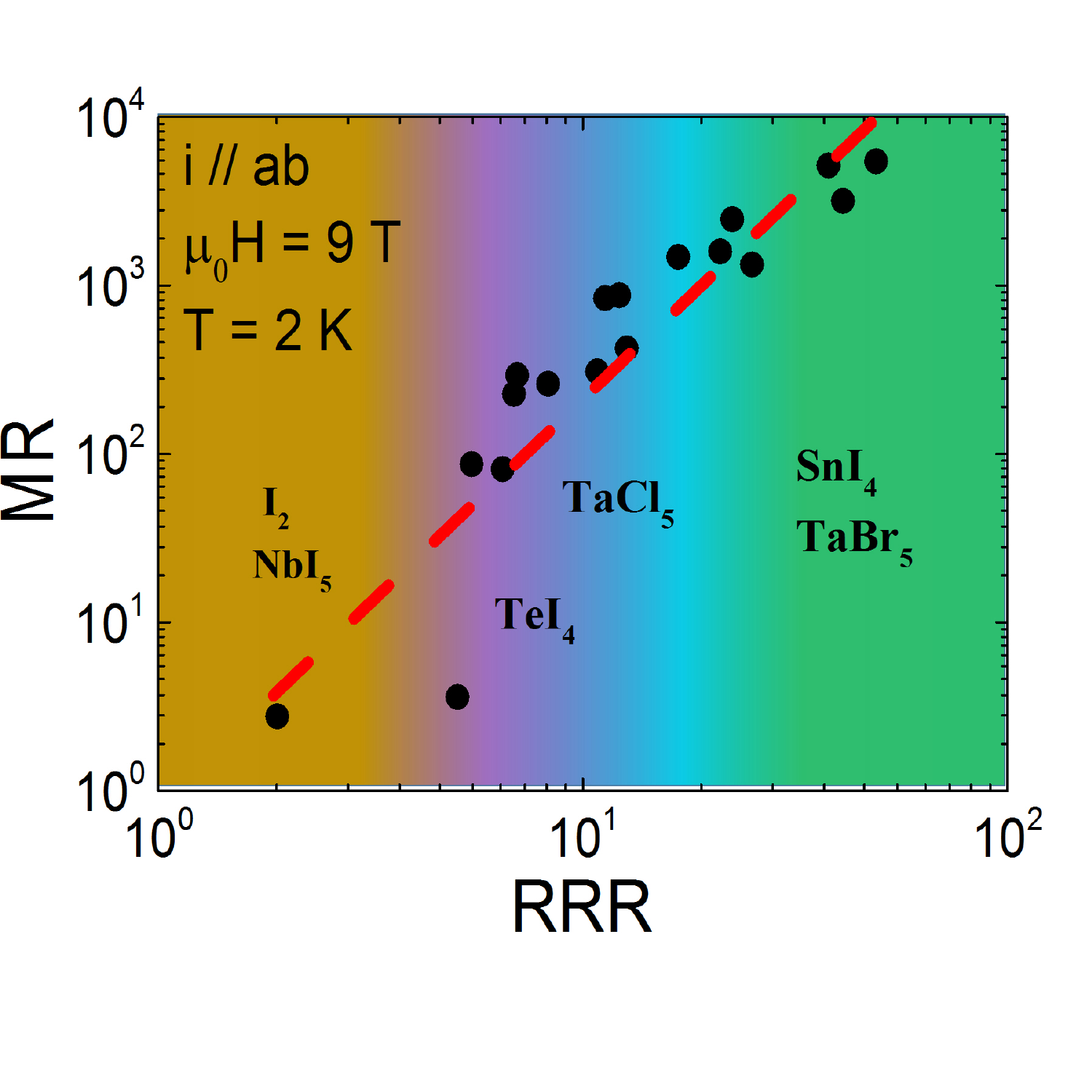}
  \caption{ \textbf{The MR and RRR for various samples of TaAs.}}
  \label{1}
  \end{center}
\end{figure}

All the temperature-dependent and field-dependent resistivity measurements were performed in a Quantum Design physical property measurement system (PPMS-9). A standard four-probe method for resistivity measurements was adopted with employing silver paste contacts on the samples. All the magneto-transport measurements were carried out from field -$H$ to $H$, and then $\rho_{xx}$ and $\rho_{yx}$ were calculated using the formulas $\rho_{xx}(H)$ = ($\rho_{xx}(H)$ $+$ $\rho_{xx}(-H)$)/2 and $\rho_{yx}(H)$ = ($\rho_{yx}(H)$ $-$ $\rho_{yx}(-H)$)/2, which eliminates the nonsymmetrical effect of the contacts.
\begin{figure}[!h]
  \includegraphics[clip, width=0.45\textwidth]{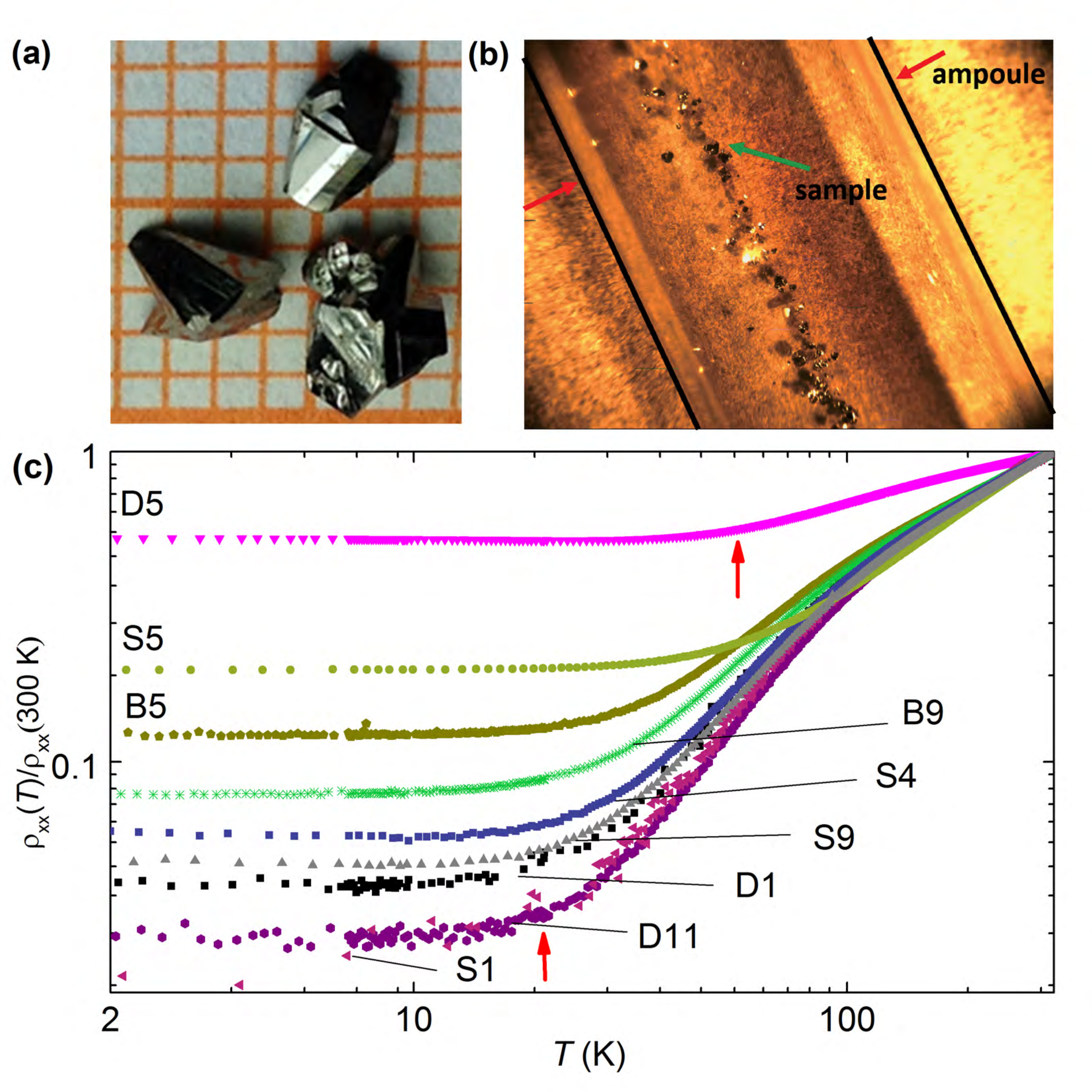}\\
  \caption{\textbf {Samples and temperature-dependent resistivity for TaAs.}  (a) A photo shows several big TaAs single crystals, the red mesh grids are 1$\times$1 mm. (b) A photo of as-grown single crystals of TaAs in an ampoule. (c) The temperature dependent resistivity of single crystalline TaAs at zero magnetic field shows a metallic profile with a $\mathrm{RRR}$ ranging from 2 to 49 for different samples. The crossovers to plateaus indicated by red arrows on the double-logarithmic plot obviously shift towards low temperatures as the $\mathrm{RRR}$ increases.}
  \label{2}
\end{figure}

\section{\romannumeral3. experiment}
We show the electrical transport properties for sample S1 in detail which has the largest RRR value. The data for other representative samples with smaller RRR are shown for comparison as well. Figure 3(a) and (b) show the temperature dependent resistivity in different magnetic fields and the $\mathrm{MR}$ at different temperatures, respectively. When a low magnetic field (0.3 T) is applied, the temperature dependent resistivity changes to an insulating profile and emerges with a plateau at low temperatures (Fig. 3(a)). The metal-to-insulator-like `transition' in magnetic field is commonly observed in semimetals such as bismuth and graphite \cite{MITchinese}. We noticed that this behavior is not like that of Cd$_3$As$_2$ \cite{liang2014ultrahigh}, in which the `turn-on' temperatures (defined as the temperatures when d$\rho$/d$T$ changes the sign) in the same magnetic fields are much lower \cite{liang2014ultrahigh}. The resistivity plateaus at low temperatures in magnetic fields are commonly observed in semimetals as well \cite{ali_WTe2_2014}.
\begin{figure}[!h]
  \includegraphics[clip, width=0.5\textwidth]{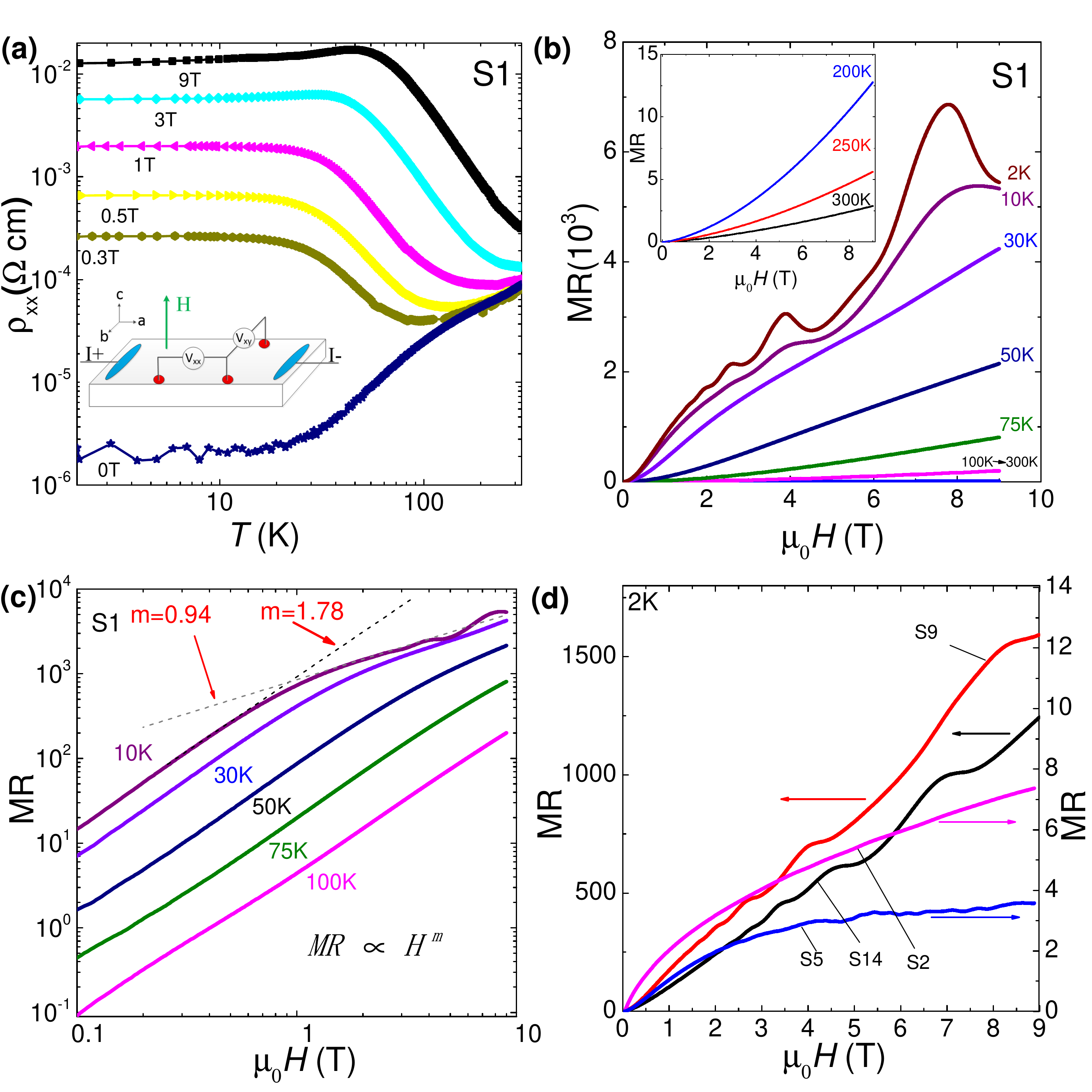}\\
  \caption{\textbf {Large MR for TaAs.}  (a) The temperature dependent resistivity in different magnetic fields. The experimental setup is shown in a sketch as an inset. (b) MR at different temperatures. Inset: MR at 200 K, 250 K and 300 K. (c) A double-logarithmic plot of MR from 10 K to 100 K. The two dashed fitting lines show the different slopes of MR at low and high fields (m = 1.78 in low fields and m = 0.94 in high fields). (d) The MR at 2 K for the samples S2, S5, S9 and S14.}
  \label{3}
  \end{figure}
Large $\mathrm{MR}$ has been observed in various semimetals including the TaAs family \cite{liang2014ultrahigh,Mun_PtSn4_2012,wang2014anisotropic,PdCoO,ali_WTe2_2014},
 but the power law varies from linear to parabolic for different compounds, and the mechanism of the large MR underneath is still under debate. Figure 3(c) is a logarithmic plot for the field dependence of the $\mathrm{MR}$ at different temperatures for sample S1. The $\mathrm{MR}$ at low temperatures changes from a parabolic to a linear dependence with a crossover manner. The crossover field changes from 0.5 T at 2 K to 6 T at 75 K.

Figure 3(d) shows the MR for several representative samples at 2 K. For the samples S2 and S5 with RRR about 5, their MR at 2 K is two orders of magnitude less than those of S9 and S14. It is noteworthy that the MR for S2 and S5 do not show linear-field dependence but intends to be saturated in a moderate field. We observed clear SdH oscillations at 2 K even for the sample with RRR less than 5. The dependence of the MR and RRR is discussed in the following part.

Figure 4 shows that the Hall resistivity ($\rho_{yx}$) above 150 K for sample S1 is positive, linearly dependent on the magnetic field up to 9 T. When the temperature is below 100 K, a large, negative field-dependent signal occurs in high magnetic field. The Hall signal is dominated by strong SdH oscillations at 2 K, while the non-oscillatory part shows a large and negative linear-field dependence.
Figure 4(b) and (c) show $\rho_{yx}$ for different samples at 300 K and 2 K, respectively. All the samples show similar magnitude of positive Hall resistivity at high temperatures while $\rho_{yx}$ for different samples at low temperature are distinct. For the samples with large RRR and MR (such as S9 and S1), their $\rho_{yx}$ is large negative with clear SdH oscillations at 2 K. Sample S5 with RRR = 5 shows positive $\rho_{yx}$ at 2 K while the SdH oscillations are very weak.

\begin{figure}[!h]
  \includegraphics[clip, width=0.4\textwidth]{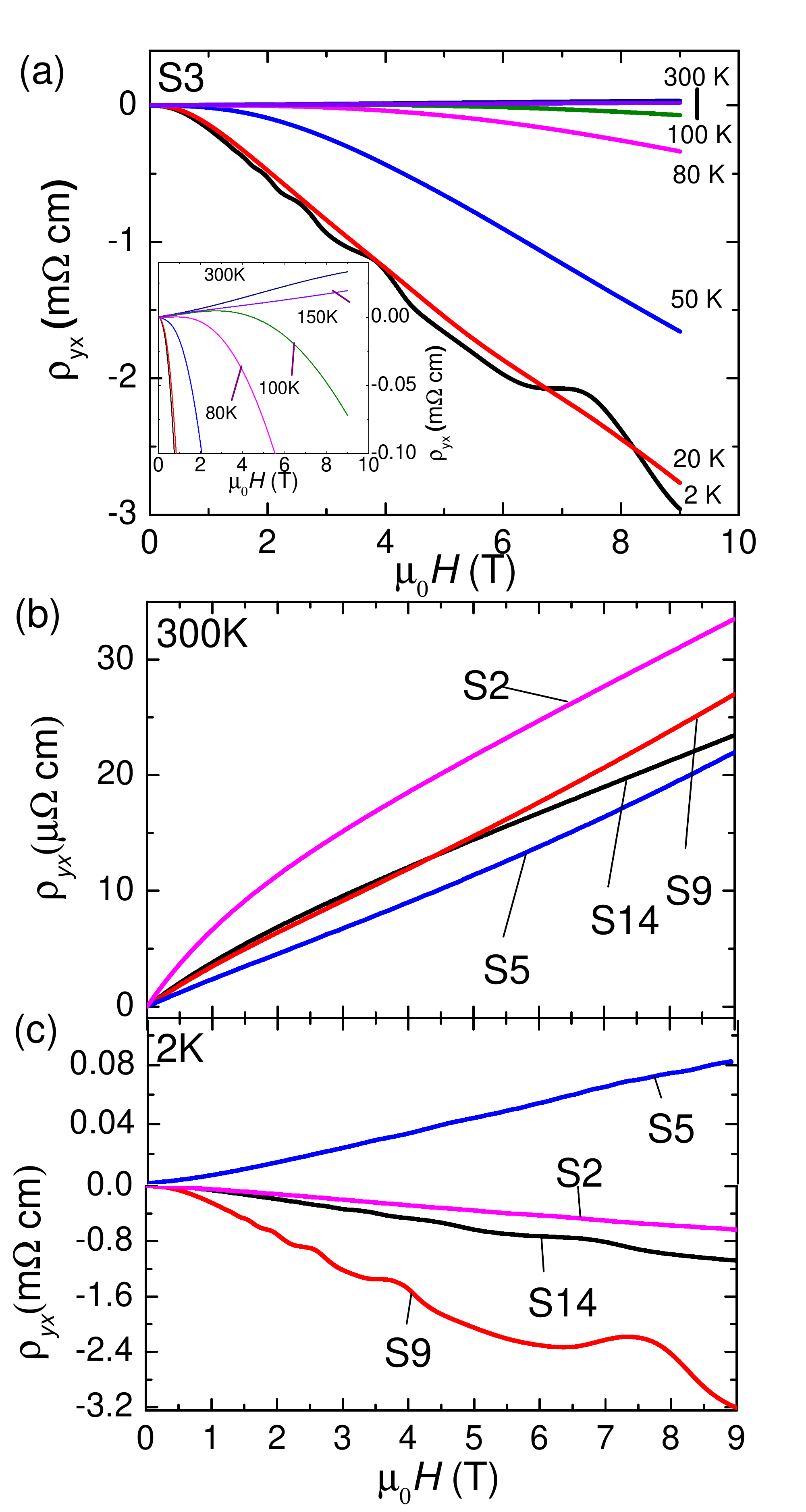}\\
 \caption{\textbf {The Hall effect for TaAs.} (a) The Hall resistivity versus magnetic fields from 2 to 300 K for sample S3. Strong SdH oscillations were observed at 2 K. Inset: The Hall resistivity at high temperature. (b) and (c) The Hall resistivity for samples S2, S5, S9 and S14 at 300 K and 2 K, respectively.
}
\label{4}
\end{figure}

\section{\romannumeral4. Data analysis}

Band structure calculations, angle-resolved photoemission spectroscopy (ARPES) and electrical transport experiments have revealed that TaAs is a semimetal possessing multi-conductive channels \cite{TaAs_ABJ_SJ}. We adopt a two-band model based on the Boltzmann equation to analyze the electrical transport data. In this model, the longitudinal conductivity reads as
\cite{hall_alloy}
\begin{equation}\label{1}
 \sigma_{xx}=en_h\mu_h\frac{1}{1+(\mu_hB)^2}+en_e\mu_e\frac{1}{1+(\mu_eB)^2},
\end{equation}
and the Hall conductivity tensor reads as
 \cite{hall_alloy}
\begin{equation}\label{2}
 \sigma_{xy}=[n_h\mu_h^2\frac{1}{1+(\mu_hB)^2}-n_e\mu_e^2\frac{1}{1+(\mu_eB)^2}]eB,
\end{equation}
where $n_e$ ($n_h$) and $\mu_e$ ($\mu_h$) denote the carrier concentrations and mobilities for the electrons (holes), respectively.

The analysis of the Hall data is based on the two-band model we outline above. We applied two constraints for the four free parameters in the formula in the fitting process \cite{ando2013topological}. They are the zero-field resistivity and the Hall resistivity in the large B-field limit. In high field, the Hall resistivity reads $\rho_{yx}=\frac{B}{e}\cdot\frac{1}{n_e-n_h}$ and the value of $n_e-n_h$ can be estimated by a linear fitting of $\rho_{yx}$. The procedure of the data analysis is standard in previous works \cite{ando2013topological}. The fittings for sample S1 at 2 K and 100 K are shown in Fig. 5. The $n_e$ and $n_h$ are well-compensated below 75 K \cite{TaAs_ABJ_SJ}. The parameters of the fittings for different samples are summarized in Table \uppercase\expandafter{\romannumeral2}.
\begin{figure}[!h]
  \includegraphics[clip, width=0.4\textwidth]{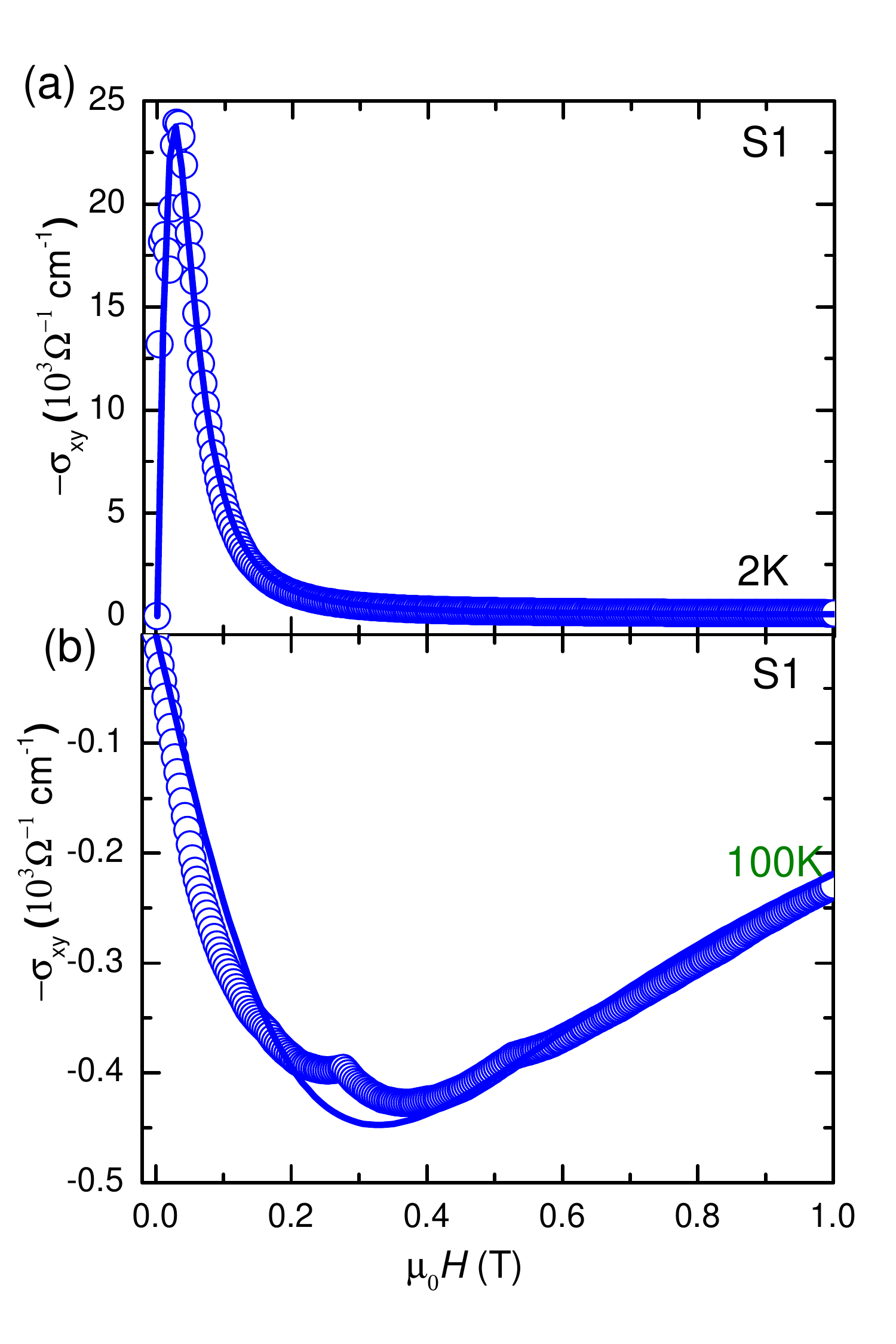}\\
 \caption{\textbf{Two-band model fitting for sample S1 at 2 K (a) and 100 K (b) (The kinks are caused by errors during data analyses in low field).} }
 \label{5}
\end{figure}

\begin{table}[h!]
\caption{Fitting results of the Hall data (2 K and 300 K) for representative samples S1, S2, S5, S9 and S14. $n$ and $\mu$ are obtained from the two-band model.
}
\label{tab2}
\begin{ruledtabular}
\begin{tabular}{lccccc}
       & $n_e$(2K) & $n_h$(2K) & $\mu_e$(2K) & $\mu_h$(2K) & $n_h$(300K) \\
unit & $10^{17}\mathrm{cm^{-3}}$ & $10^{17}\mathrm{cm^{-3}}$ & $\mathrm{cm^2(V{s})^{-1}}$ & $\mathrm{cm^2(V{s})^{-1}}$ & $10^{18}\mathrm{cm^{-3}}$ \\
\hline
S1 & $7.1$ & $4.6$ & $4.8\times10^{5}$ & $5\times10^{4}$ & $13$\\
S2 & $41$ & $10$ & $2.5\times10^{3}$ & $\textendash$ & $87$\\
S5 & $\textendash$ & $9.1$ & $\textendash$ & $4.5\times10^{4}$ & $15$\\
S9 & $6.4$ & $1.0$ & $1.5\times10^{4}$ & $\textendash$ & $33$\\
S14 & $130$ & $90$ & $2.1\times10^{4}$ & $\textendash$ & $\textendash$\\
\end{tabular}
\end{ruledtabular}
\end{table}

In order to understand the difference of the transport properties for various samples of TaAs, we analyzed the SdH oscillations in their field-dependent resistivity at different temperatures. Figure 6(a) shows the oscillatory components $\Delta\rho_{xx}$ of several representative samples at 2 K. We adopted the Lifshitz-Kosevich (L-K) formula of $\rho_{xx}$ for a 3D system \cite{murakawa2013detection}:
\begin{equation}\label{3}
 \rho_{xx}={\rho_0}[1+A(B,T)cos2\pi(F/B+\gamma)],
\end{equation}
where A(B,T) is expressed as
\begin{equation}\label{4}
 A(B,T)\propto exp(-2\pi^2k_BT_D/\hbar\omega_c)
 \frac{\lambda(T)}{sinh(\lambda(T))},
\end{equation}
where $\rho_0$ is the non-oscillatory part of the resistivity, $A(B,T)$ is the amplitude of SdH oscillations, $B$ is the magnetic field, $\gamma$ is the Onsager phase, $k_B$ is  Boltzmann's constant, and $T_D$ is the Dingle temperature. The cyclotron frequency is $\omega_c$ = e$B$/$m_{cyc}$ and $\lambda(T)$ = $2\pi^2k_BT/\hbar\omega_c$. The frequency of the oscillations is $\mathrm{F=\frac{\hbar}{2\pi{e}}A_F}$, where $\mathrm{A_F}$ is the extremal cross-sectional area of the FS associated with the LL index n, $e$ is the electron charge, and $h$ is the Planck's constant($\hbar = h/{2\pi}$). The Landau fan diagram applied here show the Berry phase for the electrons of pocket close to $\pi$ \cite{TaAs_ABJ_SJ}. The cyclotron mass was obtained by fitting to temperature-dependent SdH amplitude damping by L-K formula (Fig. 6(b)).
In order to obtain the quantum lifetime $\tau_Q$, the Dingle temperature ($T_D$) is fitted by using Eq. (4) for representative samples (Fig. 6(c)). The values of $\tau_Q$ is calculated through $\tau_Q$ = $\frac{\hbar}{2\pi{k_B}T_D}$.
All the parameters are summarized in Table \uppercase\expandafter{\romannumeral3}.

\begin{figure}[h!]
  \begin{center}
  \includegraphics[clip, width=0.5\textwidth]{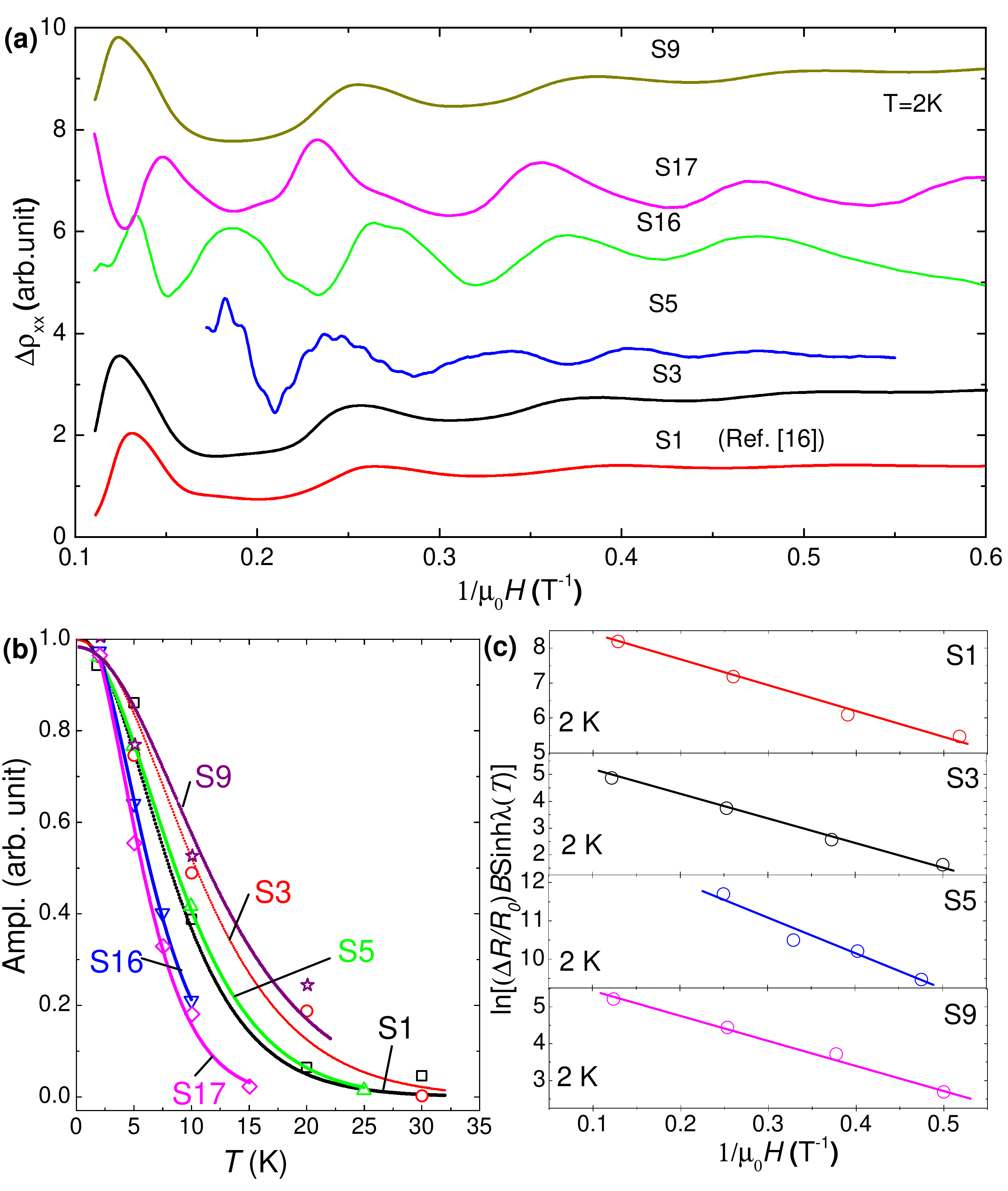}
  \caption{\textbf {Analyses of the SdH oscillations for different samples of TaAs.} (a) The oscillatory parts of $\rho_{xx}$ for representative samples at 2 K. (b) The temperature dependent amplitude of the SdH oscillations for the cyclotron mass fitting. (c) Dingle temperature plots for different samples at 2 K. }
  \end{center}
  \label{6}
\end{figure}

\section{\romannumeral5. Discussion}


\begin{figure}[h!]
\begin{center}
  \includegraphics[clip, width=0.4\textwidth]{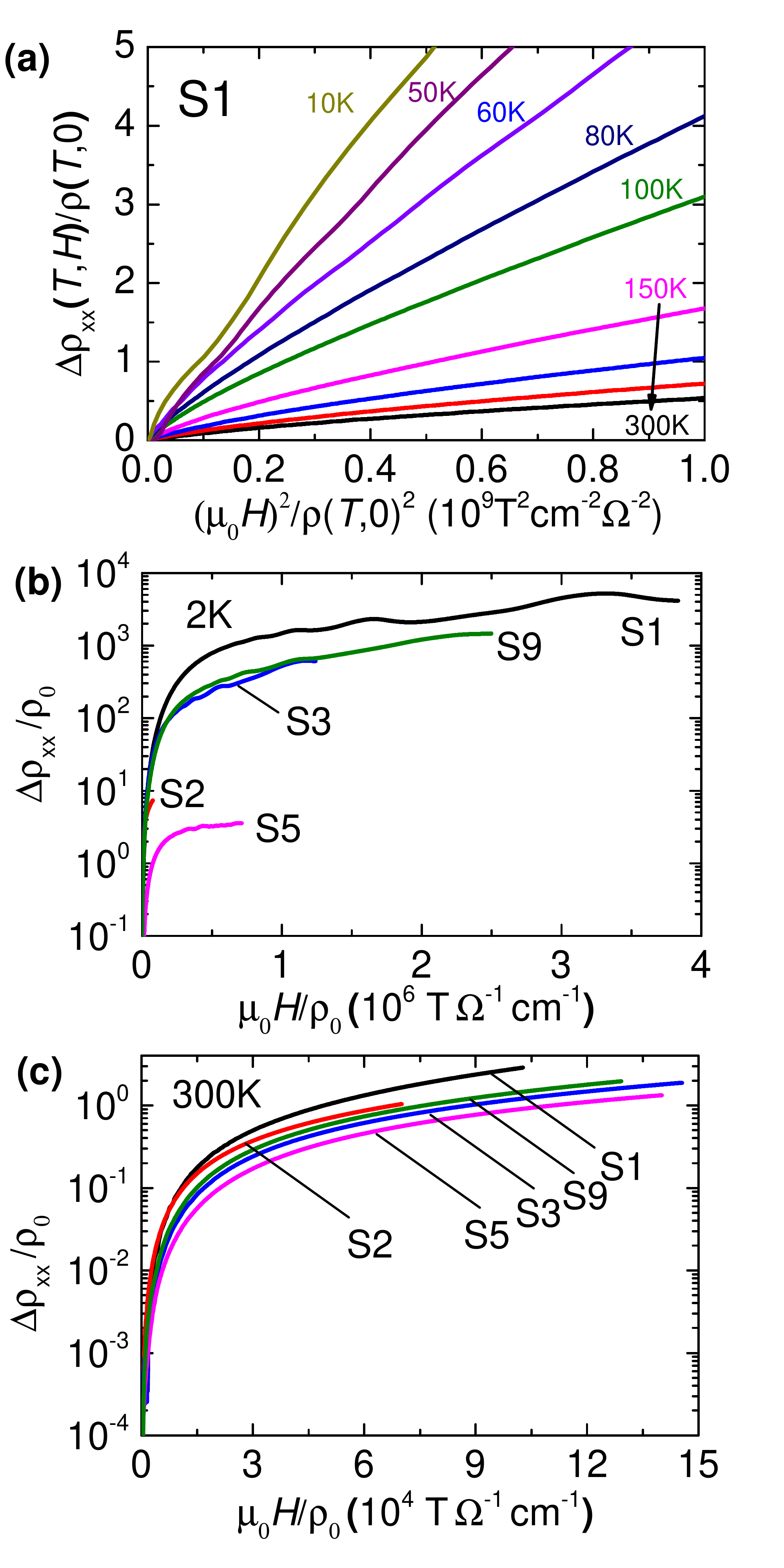}\\
 \caption{\textbf{Kohler's analysis for TaAs.}(a) The Kohler's plot of $\Delta\rho_{xx}/\rho(T,0)$ vs. $(B/\rho(T,0))^2$ for MR of sample S1 from 10 K to 300 K. (b) and (c) Kohler's plots for different samples at 2 K and 300 K, respectively.  }
  \end{center}
  \label{7}
\end{figure}
In the framework of the two-band model. The MR can be written as \cite{ziman1972principles}
\begin{equation}\label{5}
 \mathrm{MR}\equiv\frac{\Delta\rho}{\rho_0}=
 \frac{\sigma_e\sigma_h(\mu_e-\mu_h)^2B^2}{(\sigma_e+\sigma_h)^2+B^2(\mu_e\sigma_h+\mu_h\sigma_e)^2},
\end{equation}
where $\sigma_e$ ($\mu_e$) and $\sigma_h$ ($\mu_h$) are the conductivities (mobilities) associated with electrons and holes, respectively. The Drude conductivity writes as $\sigma_i$ = $n_ie^2\tau_i/m_i^*$ (i = e or h), and the corresponding mobility appears as $\mu_i$ = $e\tau_i/m_i^*$. This MR follows a quadratic $B$ dependence in low field, and is saturated in high field. If the electrons and holes share the same life time $\tau$ ($\tau_e$ = $\tau_h$), the MR will scale as a function of $\tau{B}$ and then follows a Kohler's law which states as
\begin{equation}\label{6}
 \frac{\Delta\rho_{xx}(T,H)}{\rho_{xx}(T,0)}=\mathrm{F}(\frac{H}{\rho_{xx}(T,0)}),
\end{equation}
where the replacement of $\tau$ by $\rho_{xx}(0)^{-1}$ is due to that $\tau$ is inversely proportional to $\rho_{xx}(0)$. However the Kohler's law does not constrain itself on the two-band model, its physics only underlies in the single lifetime $\tau$ assumption \cite{ziman_elePhon_1969}.

Figure 7(a) shows the Kohler's plot for sample S1 at different temperatures. The Kohler's plot is such assemblies of MR data in one sample obtained at different temperatures, or MR data of different specimens of the same compound at identical temperature versus $\frac{B}{\rho_{xx}(T,0)}$. The data all fall on a same curve in Kohler's framework.
Here the $\mathrm{MR}$ for S1 violates Kohler's rule over the whole fields. Figure 7(b) and (c) show the Kohler's plots for different samples at 2 K and 300 K, respectively. The MR for different samples severely violates the Kohler's rule at 2 K, but the MR at 300 K approximately fall on a same curve. We point out that the Hall resistivity for different samples is similar and positive at 300 K but distinct at 2 K.

\begin{figure}[h!]
  \begin{center}
  \includegraphics[clip, width=0.45\textwidth]{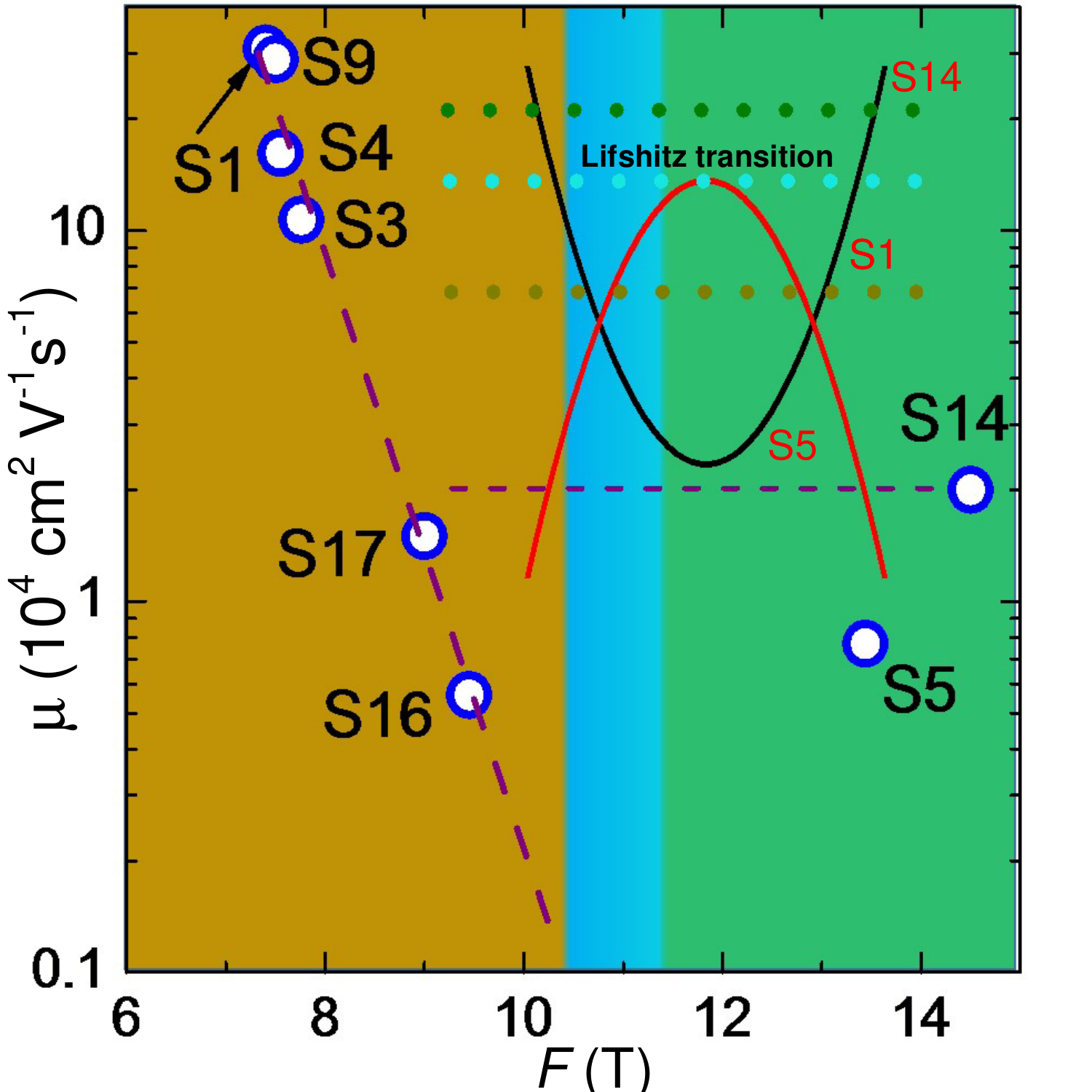}
  \caption{ \textbf{The mobility versus SdH frequency for various samples of TaAs.} The inset shows the locations of the FLs in different samples.}
  \end{center}
  \label{8}
\end{figure}

We realize that the MR at different temperatures doesn't follow the rule in Kohler's framework. The breakdown of Kohler's law can be triggered by multiple reasons \cite{mrinmetals}, while here we emphasize three plausible reasons under the background of semimetals. The first reason for breakdown of Kohler's law is the violation of the assumption of single $\tau$, saying that there are more than one $\tau$ with different values in a system. The second reason for breakdown of Kohler's law is the temperature variation of the constants of Eq. (5). In semimetals, the carrier concentration ($n$) of each pocket changes at different temperatures due to the thermal population effect. This effect can be treated as breaking of Kohler's law, but the single lifetime $\tau$ assumption may not be violated \cite{HHWKohler}. The third reason comes from the phonon scattering which is dominant at high temperature. This phonon dominated scattering will alter the scattering pattern, and then the single $\tau$ assumption is violated.  The breakdown of Kohler's law in TaAs over the whole temperature regime (Fig. 7(a)) is pertinent to all the three reasons by considering its semimetal-like electronic structures. The violation of Kohler's law for different samples at 2 K seems to be obvious because the carrier's type and concentration are different.

The information of the FL and band structure of TaAs have been obtained from the SdH oscillations, ARPES experiments and first-principle calculations \cite{TaAs_ABJ_SJ}. The main frequency of the SdH oscillations for TaAs comes from the electron pockets near the W1 nodes \cite{TaAs_ABJ_SJ}. We plot the mobility of the carriers near W1 ($\mu$) at 2 K which are obtained from the fitting by the two-band model and the main frequencies ($F$) for different samples in Fig. 8. The mobilities show a dip-like feature with respect to $F$ when 10 T $<$ $F$ $<$ 12 T. When $F$ is less than 10 T, the mobilities show exponential-like enhancement with the decreasing of $F$. When the FL is close to W1, the mobility is significantly enhanced. The external cross sectional area about 10 to 12 T corresponds to the Lifshitz transition point \cite{fourcompound} which has a flat top of the inverse band. This flat band may induce significant scattering of the electrons and suppress the mobility.

\begin{table}[!h]
\begin{flushleft}
\caption{\label{table4} Summarized transport parameters for various samples of TaAs. }
\begin{ruledtabular}
\begin{tabular}[t]{lllllllll}
$TaAs$  & E$_F$ & $\mu$(W1) & k$_F$ & m$_c$ & T$_D$ & $\tau_Q$ & $\tau_{tr}$ & $\tau_{tr}$/$\tau_Q$ \\
   & meV & m/(vs) & $\AA^{-1}$ & m$_e$ & K & s & s & \\
Sample\#   &  &  &  &  &  & 10$^{-12}$ & 10$^{-12}$ & \\
\colrule
S1 & 11.5 & 31 & 0.015 & 0.15 & 3.6 & 0.34 & 26 & 79 \\
S3 & 16.1 & 11 & 0.015 & 0.11 & 5.4 & 0.23 & 6.7 & 30\\
S5 & -45.4 & 0.77 & 0.021 & 0.07 & 9.8 & 0.12 & 0.27 & 2\\
S9 & 16.8 & 0.56 & 0.015 & 0.1 & 3.8 & 0.32 & 17 & 54\\
S14 & 24.1 & 2.1 & 0.021 & 0.14 & $\textendash$ & $\textendash$ & $\textendash$  & $\textendash$ \\
S16 & 13.6 & 1.5 & 0.017 & 0.16 & 1.1 & 1.1 & 0.52 & 0.5\\
S17 & 18.2 & 29 & 0.017 & 0.11 & 1.9 & 0.66 & 0.98 & 1.5\\
\end{tabular}
\end{ruledtabular}
\end{flushleft}
\end{table}

\begin{figure}[h!]
  \begin{center}
  \includegraphics[clip, width=0.4\textwidth]{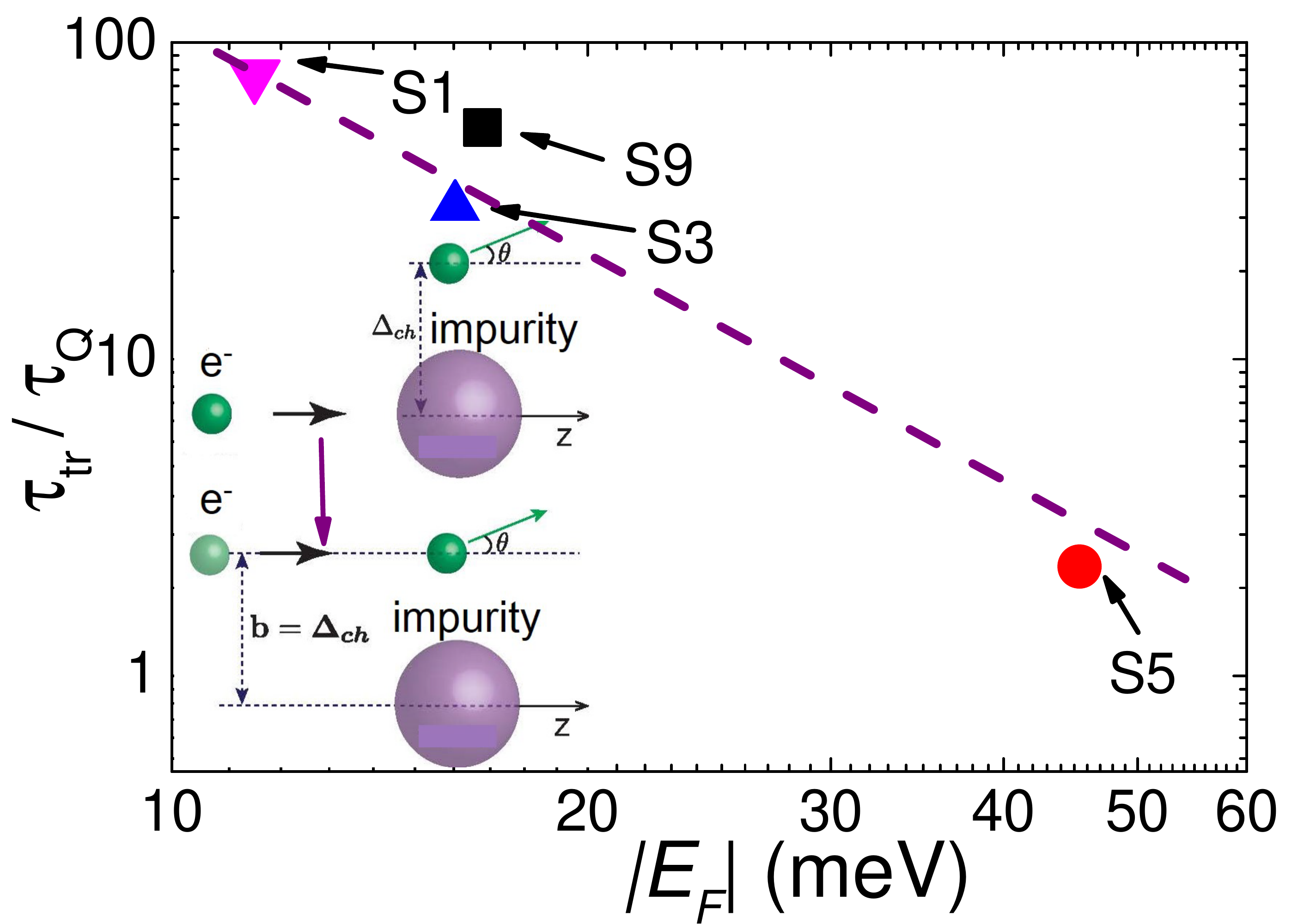}
  \caption{The ratio factor of $\tau_{tr}$/$\tau_Q$ versus Fermi energy for different samples of TaAs. Inset: A simple sketch of showing the underlying side-jump mechanism for the ratio factor enhancement, where $\textbf{b}$ is the impact parameter, $\theta$ is outgoing direction of the scattered electrons and $\Delta_{ch}$ is the chirality-protected shift
  which appears in the wave-packet scattering process. }
  \end{center}
  \label{9}
\end{figure}

The large enhancement of the mobility can be understood by analyzing the scattering process where the FL is close to the W1 node. By taking sample S1 for example, We calculated quantum life time $\tau_Q$ = $\frac{\hbar}{{2\pi}k_BT_D}$ = $3.4\times10^{-13}$ $\mathrm{s}$ by fitting the Dingle temperature $T_D$ ($T_D$ = 3.6 K). The transport life time $\tau_{tr}$ is estimated from the expression $\tau_{tr}$ =  $\frac{\mu_e{\hbar}k_F}{ev_F}$ = $2.6\times10^{-11}$ $\mathrm{s}$. It is well-known that $\tau_{tr}$ measures backscattering processes that relax the current while $\tau_Q$ is sensitive to all processes that broaden the LLs. The large ratio $R_\tau\equiv\tau_{tr}/\tau_Q$ = 79 for sample S1 indicates that the small-angle scatterings are dominant while the backscatterings are strongly protected at low temperature. The same transport parameters are also obtained in other samples, and summarized in Table \uppercase\expandafter{\romannumeral3}. We select the values of the samples from a same batch, and plot them in a double-logarithmic plot in Fig. 9. We can see an evident enhancement of $\tau_{tr}$/$\tau_Q$ when the FL approaches the W1 node. A recent theoretical work unveiled the underlying mechanism of the protection against the backscattering \cite{jiang2016chiral}. In a Weyl/Dirac system, the electrons will gain an extra side momentum ($\Delta_{ch}$) to side jump because of the definite chirality. This backscattering protection effect is enhanced when the FL approaches the Weyl/Dirac point \cite{jiang2016chiral}.

Our analyses of the electrical transport properties for various samples of Weyl semimetal TaAs reveal that the mobility and MR are indeed strongly dependent on the samples' qualities. Particularly the positions of the FL strongly affect the scattering process and mobilities. Fine tuning the FL of the Weyl semimetal TaAs family is crucial for exploring the unique properties of the Weyl quasiparticles.

\begin{acknowledgments}
We thank Yuan Li and Ji Feng for using their instruments.
S. Jia is supported by National Basic Research Program of China (Grant
Nos. 2013CB921901 and 2014CB239302). C.L. Zhang treats his research work as precious homework from Shui-Fen Fan.
\end{acknowledgments}

\end{document}